\documentclass[aps,preprint]{revtex4-1}

\usepackage{graphics}
\usepackage{graphicx}
\usepackage{url}

\begin{document}

\title{Urban Skylines: building heights and  \\ shapes as measures of city size}

\author{
Markus Schl\"apfer$^{1}$, Joey Lee$^{2}$ and  Lu\'is M. A. Bettencourt$^{1}$}

\address{$^{1}$Santa Fe Institute, Santa Fe, NM 87501, USA\\
$^{2}$University of British Columbia, Department of Geography, Vancouver, V6T 1Z2, Canada}

\begin{abstract}
The shape of buildings plays a critical role in the energy efficiency, lifestyles, land use and infrastructure systems of cities. Thus, as most of the world's cities continue to grow and develop, understanding the interplay between the characteristics of urban environments and the built form of cities is essential to achieve local and global sustainability goals. Here, we compile and analyze the most extensive data set of building shapes to date, covering more than 4.8 million individual buildings across several major cities in North America. We show that average building height increases systematically with city size and follows theoretical predictions derived from urban scaling theory. We also study the allometric relationship between surface area and volume of buildings in terms of characteristic shape parameters. This allows us to demonstrate that the reported trend towards higher (and more voluminous) buildings effectively decreases the average surface-to-volume ratio, suggesting potentially significant energy savings with growing city size. At the same time, however, the surface-to-volume ratio increases in the downtown cores of large cities, due to shape effects and specifically to the proliferation of tall, needlelike buildings. Thus, the issue of changes in building shapes with city size and associated energy management problem is highly heterogeneous. It requires a systematic approach that includes the factors that drive the form of built environments, entangling physical, infrastructural and socioeconomic aspects of cities.  
\end{abstract}

\maketitle

\section{Introduction}

Nothing is quite as emblematic of a city as its skyline. Almost all large cities have extraordinary buildings and their shapes and profiles act together as a sort of identifier of particular places and times~\cite{Heath_2000}. This phenomenon seems to repeat itself at different scales, with regional urban centers also developing taller and denser downtowns than smaller surrounding towns. Thus, it is interesting to ask how much we can tell about a city from its skyline.   

Beyond the strong first impression that a city's skyline can produce lie a number of functional attributes of buildings that are essential for how cities operate. These include the relationships between building heights and associated energy consumption, transportation needs and patterns of land use. Very dense developed cities, such as Hong Kong or parts of Manhattan, strongly concentrate energy use in space due to high employment and population densities housed in very tall buildings. Such patterns of densification serve socioeconomic purposes but also lead to challenges and opportunities to the sustainable development of cities. In New York City, for instance, over two thirds of the energy is used in buildings, mainly for space cooling, heating, lighting and electricity driven appliances~\cite{Howard_2012}. With similarly high fractions of energy use in many modern cities~\cite{IEA_2013}, buildings are becoming the focus in the effort to understand and mitigate the negative consequences of urban energy use and associated pollution and greenhouse-gas (GHG) emissions~\cite{PerezLombard_2008, Parshall_2010, Martani_2012}. This is particularly important in developing cities, where the expansion of urban built fabric will exceed all that we have built worldwide so far. Many of these urban areas will need to grow in height to house and employ growing population sizes, as is clearly visible in the changing skylines of cities in China or India today. What are the expected patterns of growth in building heights for such cities? And how may we seize opportunities arising from the understanding of such patterns to develop more sustainable megacities?

Extensive research has focused on buildings as self-defined entities~\cite{PerezLombard_2008} or on selected and rather small areas within a particular city~\cite{Ratti_2005}. These results have demonstrated the impact of building design on the resulting thermal load. At the urban scale, however, comparatively little has been done so far, mainly due to the lack of detailed city-wide statistics of building morphologies.  Batty, Steadman and collaborators, on the basis of earlier work by Bon~\cite{Bon_1973} and Steadman~\cite{Steadman_2006}, took important steps in this direction by exploring the intra-city allometry and building statistics for Greater London~\cite{Batty_2008,Steadman_2009}. Nevertheless, quantitative insights across cities of varying population sizes are needed to derive better predictions of the built form as urban areas expand and develop. Indeed, as casual observation readily attests, building shape is determined, at least on average, by the value of land, which in turn is associated with urban wealth and population density~\cite{Bettencourt_2013}. Increasing building height creates more floor space per unit of land occupied and, thus, serves as a dimension that can be expanded in order to maintain living and working spaces affordable even as land values rise. In recent years, advances in high-resolution remote sensing technology, such as airborne Light Detection and Ranging (LiDAR), have made it possible to generate detailed data on building geometries over vast geographic areas~\cite{Tooke_2014}.  At the same time, an increasing number of city governments are opening their comprehensive building data bases to the public~\cite{Batty_2012}, and collaborative user-generated mapping projects such as OpenStreetMap (OSM) have started to incorporate 3-D building geometries from a growing number of cities worldwide~\cite{Haklay_2008}. Thus, there is a clear prospect that building shapes for all cities in the world will be available in the near future and that this sort of information will become a common basis for urban science and planning policy.

Here, we compile and analyze a new large-scale data set of buildings, unprecedented in terms of geographic scope and resolution, to derive general characteristics of building shape as a function of city population size. More specifically, we first assemble data for 12 North American cities, containing the geometries of $\approx$4.8 million buildings in total. Most of these data are based on LiDAR measurements, see Material and Methods. We then use this information to develop a predictive theoretical framework for the increase in average building height as cities become larger, by extending concepts from urban scaling theory. Subsequently, we establish the allometric relations between city size and the statistics of buildings shapes and conclude by discussing some of the potential consequences for urban energy use as functions of building sizes and shapes.

\section{Results and discussion}

\begin{figure}[t!]
\centering\includegraphics[width=1\textwidth]{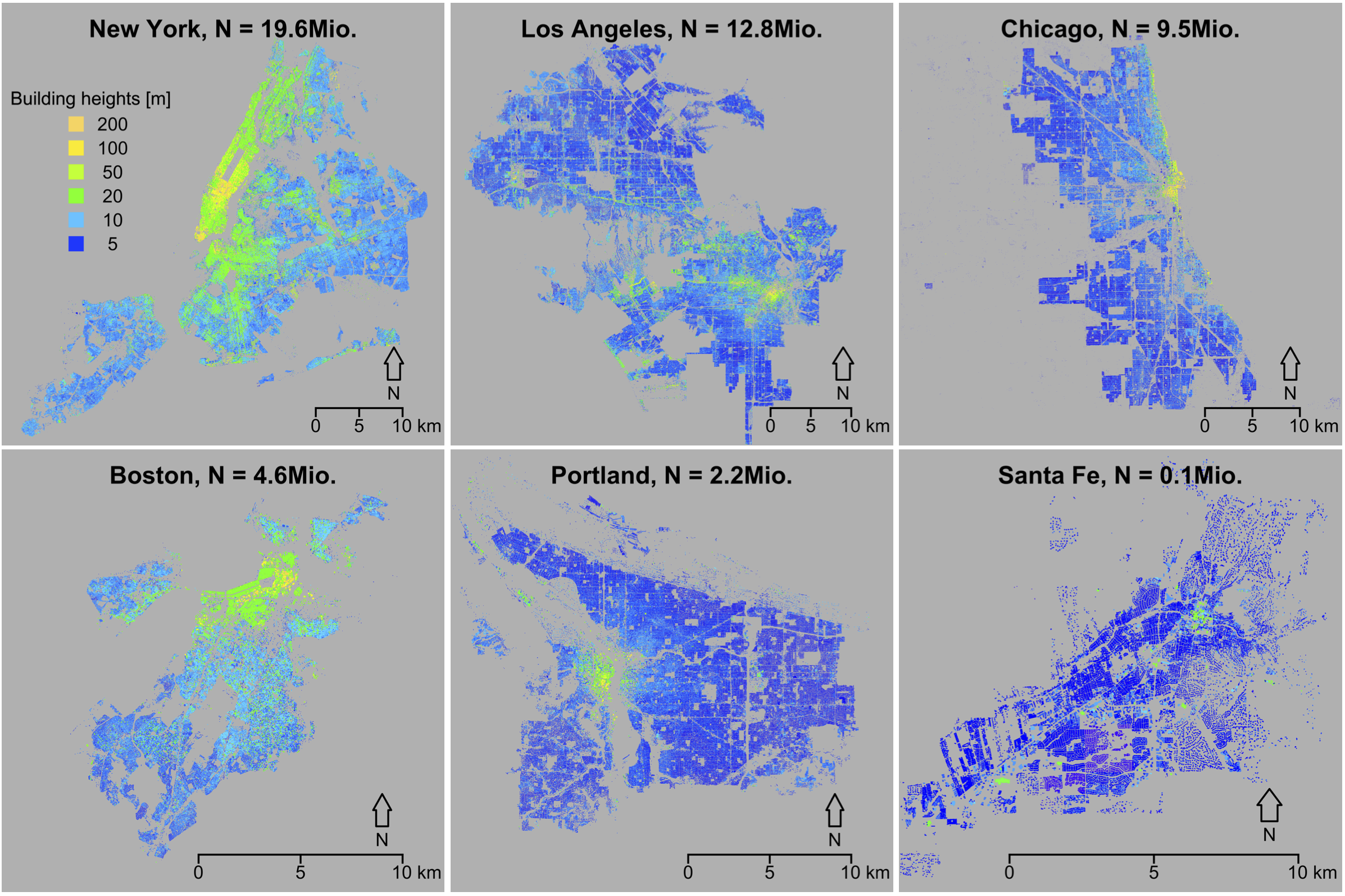}
\caption{Building height maps for 6 different US cities, spanning a wide range of population sizes. The color code denotes building heights and applies to all panels. Total population sizes, $N$,  are taken to be those of the corresponding Metropolitan Statistical Area (MSA), which is the functional socioeconomic urban area defined by the US Census Bureau (\mbox{Census 2010)}.}
\label{fig:footprint}
\end{figure}

Figure~\ref{fig:footprint} depicts the geographic distribution of building heights in a few example cities and \mbox{table~\ref{tab:statistics}} reports the summary statistics for all cities in our data set. This gives us a first impression that is in line with general intuition that larger cities have taller buildings. However, this also shows great heterogeneity in building heights across space, together with a general trend towards lower structures with a single floor ($\approx$3-5m) at the edge of the city.

Dealing with this strong spatial heterogeneity is the greatest challenge for any comparative study of building characteristics. To limit potential biases due to the highly variable geographic scope and heterogeneous spatial coverage of the different data sets (figure \ref{fig:footprint}), we first define a circle of fixed radius $d$ around each city center (defined as the location of the city hall), and subsequently compare the building statistics from within these circle areas.

\subsection{Building height statistics} 

\begin{figure}[t!]
\vspace{-1cm}
\centerline{\includegraphics[width=0.7\textwidth]{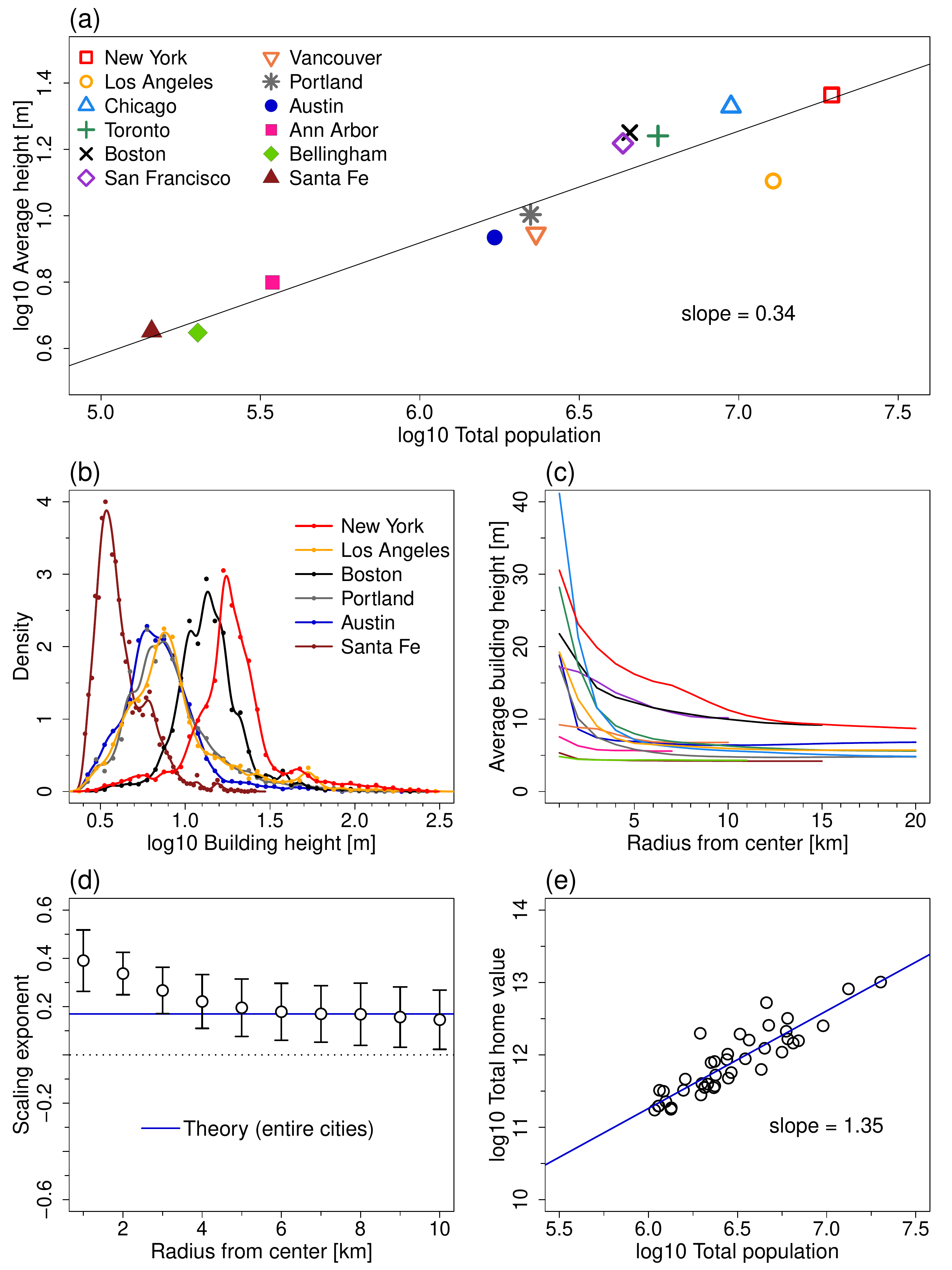}\vspace{-0.1cm}}
\label{Fig-5}
\caption{\label{fig:scaling}The variation of building height with city population size. ({\it a}) Scaling relation between average building height, $h$, and city population size, $N$, within $d=2$~km distance from the city center (location of City Hall). The best fit line has slope $0.34$, with 95\% confidence interval $[0.25,0.42]$, $R^2=0.87$. ({\it b}) Statistical distributions of individual building heights. For visual clarity, the distributions for only half of the cities in the data set are shown. The continuous lines are kernel density estimations. ({\it c}) Average building heights within radius $d$ from the city center that defines the perimeter of the analyzed area. Colors are as in panel ({\it a}). ({\it d}) Scaling exponent $\beta$ versus radius $d$ from the city center. The error bars indicate the 95\% confidence interval. The continuous line is the theoretical prediction $\beta = \delta \approx 1/6$. ({\it e}) The scaling of the value of single family homes with population size in various major US metropolitan areas (MSAs). The best fit line has slope $1.35$, with 95\% confidence interval $[1.13,1.57]$, $R^2=0.79$, being statistically indistinguishable from the prediction of equation~\ref{eq:rent}, with $1+2\delta \approx 1+1/3$.}
\end{figure}

We start by showing the behavior of average building heights, $h$, for our set of cities. Figure \ref{fig:scaling}({\it a}) depicts the values of $h$ within a rather short distance of $d=2$~km from the city center, versus population size $N$ of the corresponding metropolitan area (functional socioeconomic city). For these `downtown' areas, we observe that average building heights systematically increase with city size over more than 2 orders of magnitude (Spearman's $\rho = 0.91$, \mbox{$p$-value$<0.001$}) and that the average height $h$ can be modeled by a power law function, $h \propto N^\beta$, with scaling exponent $\beta =0.34$. The variance is surprisingly small, given a broad range of possible distortions such as building height restrictions or the specific life history of cities. A clear `outlier' is Los Angeles with a substantially lower average building height than should be expected from its population size. This result is in line with its specific spatial distribution of employment centers that have been shown to be highly dispersed over the entire urban area~\cite{Gordon_1996}. 

The underlying statistical distributions of individual building heights are depicted in \mbox{figure \ref{fig:scaling}({\it b})} for various cities. The histograms of the logarithmic $h$ values indicate heavy-tailed distributions, generalizing earlier findings~\cite{Batty_2008}. The occurrence of multiple peaks reflect the fact that buildings usually have a discrete number of floors. Taking Santa Fe, NM, as an illustrative example, we first observe a pronounced peak at $h\approx3.5$~m, reflecting the presence of a large number of single-story buildings, then a smaller peak at $h\approx6.5$~m, reflecting two-story buildings, and so on. Moreover, the multimodality also indicates that building heights are strongly clustered around different values that are typical for different areas of a city, for instance due to district-level variations in height restrictions. The general shift of the distributions towards higher values for larger cities shows that the scaling of the average height values, \mbox{figure \ref{fig:scaling}({\it a})}, is not due to the dominant effect of a few individual buildings, but results from an increase in height that characterizes most buildings in the city.

Urban economic theory suggests that the cost of space in terms of rent and consequently densities for all land uses significantly decrease with increasing distance from the city center\cite{anas1998uss}. Thus, under the assumption of such regular mono-centric patterns, the average building height should also continuously decrease with increasing cut-off distance $d$. This is indeed the case for all analyzed cities, as shown in figure~\ref{fig:scaling}({\it c}), demonstrating a clear peak of $h$ in the center. While the scaling exponent $\beta$ also decreases within short distance from the center, it remains approximately constant with a value of $\beta \approx 1/6$ (horizontal blue line) over the range of $d$ considered here, see figure \ref{fig:scaling}({\it d}). Below, we provide a theoretical framework to explain and reproduce this behavior of average building height and to parameterize changes in building size and shape as functions of city size.

\subsection{Theoretical prediction of average building height}

It is to be generally expected that larger cities have some taller buildings. There are several reasons for this, related to the concentration in space of population, infrastructure and economic activity.  Here, we explore the consequences of such densification. Specifically, by asking that average incomes stay commensurate with rents on built space we show how to derive expectation on average building heights. To do this, consider a city with population size $N$, land area $A$ and total income (or GDP) $Y$.  The built land area of the city is isomorphic with the infrastructure network~\cite{Brelsford_2015} and is denoted by $A_n$, as described in~\cite{Bettencourt_2013}. If a fraction of total income, $b$, is allocated to land rents, we obtain that the {\it average} price of land per unit area and unit time as a function of city size $N$, is 
\begin{equation}
{\rm rent}(N) = b \frac{Y}{A_n} = b \frac{G}{A_0^2}N^{2 \delta},
\label{eq:rent}
\end{equation}
where $Y(N)=G N^2/A_n$ and $A_n (N) = A_0 N^{1-\delta}$, see \cite{Bettencourt_2013}. In practice, $b$ may itself depend on city size, but we leave it here as a parameter. The most general, and simplest, expectation for the exponent is  $\delta \approx 1/6$~\cite{Bettencourt_2013}. This expectation is borne out by the value of single family homes in most major US metropolitan areas, see figure.~\ref{fig:scaling}({\it e}). These express the aggregate value of housing before stronger densification effects set in and buildings start to get tall towards city centers.

This behavior of land rents ultimately poses a problem: because of the compounded effects of greater density (less space per capita) and higher productivity and incomes (more money per capita), land rents increase with city population size {\it faster than income}, and would eventually consume and overcome all productivity gains. This is unacceptable. The way out of this dilemma relies on cities developing a systematic change in their land use as they grow. 
This adaptation can happen in practice in two ways. First, individuals and firms may adjust their lifestyles to consume less space, in line with their incomes. We can compute what such a change would imply by defining $a_f$ as the average area of {\it floor} space that an individual actually uses. Equating this to the total area and solving for the value of $a_f$, such that rents scale like per capita incomes, $y=Y/N$, we obtain
\begin{equation}
a_f \cdot {\rm rent} = b~y  \to a_f  = b \frac{y}{\rm rent} = \frac{A_n}{N} = A_0 N^{-\delta}.
\end{equation}
This result entails a relative reduction of about $\delta \simeq 17\%$ in area per person with every doubling of population size. Although smaller spaces per capita are certainly a feature of many large cities, the ultimate expression of this logic is to be found in informal settlements or slums. It is estimated, for example, that the population density of the Lower East Side of Manhattan in the 19$^{\rm th}$ century reached almost 150,000 people/km$^2$, the highest population density ever in the US, and likely to never again be repeated. 

Second, to avoid having to use less space per capita, cities may simply produce more floor space per unit of land. This requires taller buildings. It is readily observable that as cities grow and develop they tend to use more of their third spatial dimension, by burying much of their infrastructure and by building taller residential and commercial buildings.  The growth in height of a building typically incurs additional costs and is regulated through zoning. Costs are the result of both necessary changes in direct construction and in a relative reduction in usable space as buildings grow taller due to structural requirements to support more weight and building infrastructure~\cite{Costing_Buildings}. These include stairwells, elevator shafts, cables, pipes, etc. The ratio of floor space to land area also depends on the shape of the building. Familiar examples are the growth and adaptation of elevator shaft design: in tall buildings elevators are typically built in terms of double or even triple deckers, meaning that a person wanting to reach the top of the building will change elevators 2-3 times. Many architects see it as a rule of thumb that with present technology a building taller than about 100 stories is not economically viable.  We model these effects through a dimensionless number, $0 < C(h) \leq 1$, which accounts for the relative loss of floor space as building height, $h$, increases. Nevertheless, at moderate heights it is estimated that the monetary cost of buildings is approximately linear on height, incurring no especially large penalty in costs~\cite{Costing_Buildings}.

The point about making a building taller is to compensate for competition for land by creating more floor space. To compute the expected average height of buildings in a city of population $N$ and GDP $Y$, we require that $a_f$ is a constant, independent of city size, and write the floor space per capita as
\begin{eqnarray}
a_f=\frac{A_n}{N} C(h) \frac{h}{h_0} \rightarrow h(N) = \frac{h_0 a_f N}{A_n C(h)}=\frac{h_0 a_f}{A_0 C(h)} N^\delta.
\end{eqnarray} 
The function $C(h)$ is not generally known~\cite{Costing_Buildings}. Measures of building costs often assume a linear relationship between total costs and height. Hence, if loss of space were neutral with height, at least initially, we predict that average height increases with an exponent around $\delta \simeq 1/6$, similar to built density~\cite{Angel_2011}. Given that $C(h)$ will typically decrease with height, we should expect a slightly larger exponent, particularly when considering only areas around the downtown cores where buildings tend to be tall (i.e. for small values of the radius $d$). This theoretical prediction is consistent with our empirical observations, see figure \ref{fig:scaling}({\it d}). We develop a simple model for $C(h)$ and explore some of its consequences in the electronic supplementary material.

\subsection{Building allometry and energy efficiency}

We now compute the changes in building shape across cities and discuss their general consequences for energy management. In particular, we want to estimate the surface-to-volume ratio, $r_b$, for different buildings. This is intimately connected with building allometry, that is, how different dimensions of building shapes co-vary. 

The quantity $r_b$ is important because it determines the energy efficiency of buildings: it is through its external surface that a building receives light  from the outside and maintains a gradient of temperature for climate control, expending energy in the process. There are various tools to compute the energy efficiency of buildings, such as {\it EnergyPlus} from the US Department of Energy~\cite{Crawley_2001}. For example, to compute energy use due to climate control, these tools rely ultimately on elaborations of a model for thermal diffusion through the building's walls. Such models obey a thermal equation of the type
\begin{eqnarray}
\frac{d E}{dt} = \mu \int dn ~{\vec \nabla T} .{\vec n}, \rightarrow E (\tau) = \mu \int^\tau dt \vert {\nabla T} \vert S_b= \tau \mu S_b \Delta T,
\end{eqnarray} 
where $E(\tau)$ is the energy spent over a time interval $\tau$. The vector ${\vec \nabla T}$ is the gradient in temperature across the building's walls, between the building's interior and the exterior, normal to the wall $\vec n$. The parameter $\mu$ accounts for both the efficiency of converting energy into an ambient temperature and the average thermal diffusion across the walls (and its associated efficiency of insulation). The building's total surface is given by $S_b$. In the last expression, on the right, we took the average temperature gradient over the time $\tau$, as $\Delta T$.  

The energy efficiency of the building should be evaluated on a per capita basis and not in absolute terms as large buildings may house many people. To do this we compute the expected number of people occupying a building as 
\begin{equation}
N_b = \frac{\ell^2}{a_f} C(h) \frac{h}{h_0} = \frac{V_b}{v_b},
\end{equation}
where $\ell^2$ is the footprint area. The quantity $v_b = a_f h_0$ is the average building volume per capita, which we take to be city size independent. Thus, the energy used per capita in the building is 
\begin{eqnarray}
\frac{E(\tau)}{N_b} = \tau \mu v_b  \Delta T r_b , \quad r_b=\frac{S_b}{V_b}.
\end{eqnarray}
We see that energy efficiency is related to building size and shape through the surface-to-volume ratio, $r_b$, $E / N_b \propto r_b$. The objective of allometry is to relate a building's surface area, $S_b$, to its volume, $V_b$. The simplest example is for a spherical building.  Its shape is characterized by a single length scale, which is the diameter $D$. For a sphere, we have that the building surface is $S_b= \pi D^2$, and its volume is $V_b = \frac{\pi}{6} D^3$. It follows immediately that the surface-to-volume ratio for a sphere is $r_b=\frac{S_b}{V_b} = \frac{6}{D}$.  

The sphere is special because it is the shape with the smallest surface-to-volume ratio. We therefore take the sphere as a reference when generalizing to more realistic building shapes. Our starting point is a cube with side length $D=V_b^{1/3}$. For the cube it is also the case that $r_b=\frac{6}{D}$. The next step in generalizing shapes is to consider a square cuboid shape, with footprint area $\ell^2 =A_f$ and height $h$. We now have two characteristic length scales $\ell$ and $h$, and the surface to volume ratio will be a function of both. We have that 
\begin{eqnarray}
&& S_b = 2 \ell^2 + 4 h \ell; \qquad V_b = \ell^2 h, \\
&& r_b = 2 \left( \frac{1}{h} +   \frac{2}{\ell} \right) = \frac{6}{D} \left( \frac{  x^{-2/3} + 2 x^{1/3}}{3} \right),
\label{eq:cuboid}
\end{eqnarray}
where $D=(\ell^2 h)^{1/3}$ and $x=h/\ell$ is a dimensionless shape parameter. Minimizing $r_b$ relative to $x$ at fixed $D$, one obtains the shape with the smallest surface to volume ratio: this is the most spherical shape, corresponding to $\ell=h$ or $x=1$.  For any other shape, the term in brackets involving $x$ will be larger than one. This expression is easy to generalize to buildings that have an additional length, for instance if the footprint is not well described by a square but rather resembles a long rectangle. This result and the procedure to its generalization to other shapes is presented in the electronic supplementary material.  

It follows from equation~(\ref{eq:cuboid}) that the shape of any building can be characterized by two numbers, an extensive quantity $D$ (a length) measuring the overall size of the building, and a quantity $x$, characterizing how far this shape deviates from a cube. We can now easily derive the allometric relation between $S_b$ and $V_b$, using the expressions for $r_b(D,x)$ and the fact that $D=V_b^{1/3}$, as
\begin{eqnarray}
S_b(V_b,x) = r_b(V_b,x) V_b= 6 V_b^{2/3} \left( \frac{  x^{-2/3} + 2 x^{1/3}}{3} \right).
\end{eqnarray}
We see therefore that the simplest allometric relation $S_b\sim V_b^{2/3}$ requires that shape parameters are independent of $D$. To the extent that such parameters are correlated, since taller buildings on average tend to require a larger base, such a relation will receive a correction.  

Thus, to conclude, the allometric relation for buildings and their surface to volume ratio, depend not only on their size $D$ but also, in general, on shape parameters. $D$ measures the overall size (or extensive) efficiency of a building, whereas $x$ measures shape (or intensive) efficiency. For $x<<1$ the building looks more like a planar sheet and for $x>>1$ more like a needle; the point at which $x=1$ is the most cube-like shape and the place for which one predicts the canonical allometric relation $S_b \sim V_b^{2/3}$.  This is also the shape for which buildings are most efficient in terms of climate control. 

\begin{figure}[t!]
\includegraphics[width=0.38\textwidth]{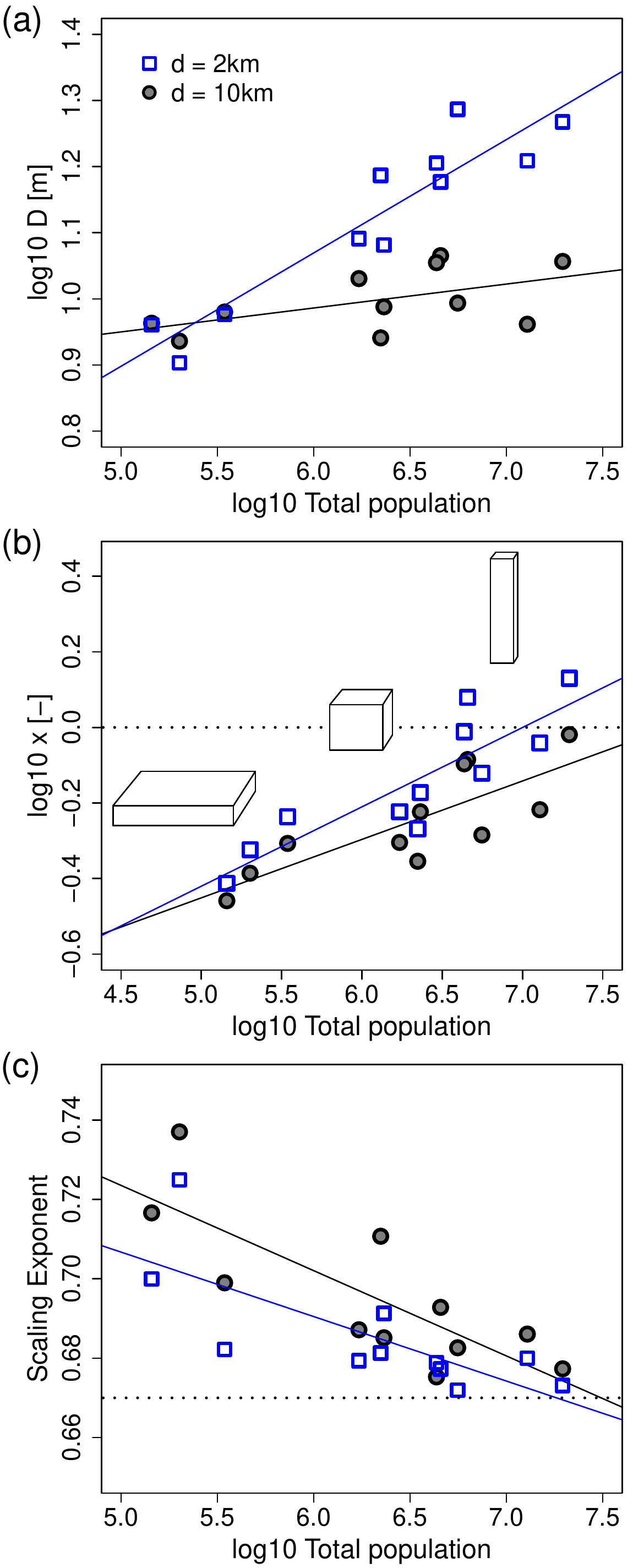}
\caption{\label{fig:distributions} The dependence of average building size and shape on city size, and resulting allometric scaling behavior. ({\it a}) Building size parameter $D$ for two different radii $d$ from the city center (without Chicago as our data did not contain building footprints). ({\it b}) Ratio of the characteristic building length scales, $x=h/\ell$. ({\it c}) Exponent of the allometric scaling relation. We observe that buildings become larger with city population size (larger $D$) and that their shapes change and become, on average, more like cubes ($x\rightarrow 1$). In this limit, but not before, the city-wide allometric scaling relation approaches its simplest form, with $ S_b \sim V_b^{2/3}$. }
\label{fig:allometry}
\end{figure}

Figure~\ref{fig:allometry} reports the variation in $D$ and $x$ with city size for two radii around the city center. The behavior of $x$ versus $D$ is shown for all cities in the electronic supplementary material. We see that building sizes do increase with city size, creating the conditions for greater energy efficiency in terms of climate control, see figure~\ref{fig:allometry}({\it a}). Shapes also change and move, on average, towards $x=1$, see figure~\ref{fig:allometry}({\it b}), making the allometric relation converge towards its simplest form for approximately `spherical' buildings, see figure~\ref{fig:allometry}({\it c}). Only for New York City and Boston we observe values of $x>1$ for $d=2$~km, indicating that the surface-to-volume ratio increases again in the downtown cores of large cities, due to the proliferation of tall, needlelike buildings. 

\section{Conclusion}

Urban skylines tell us much about the characteristics of a city. We have analyzed the shapes of about 4.8 million buildings within and across 12 cities in North America to show that, while the specific shape, location and organization of buildings vary substantially, there are a number of statistical signatures of city population size and wealth in each urban skyline. Larger cities have on average taller and bigger buildings as an adaptation to higher population and employment densities and levels of income. The shape of buildings changes on average with city size, starting out fairly flat in small cities to become increasingly cubic in larger cities, whose downtowns also feature a high prevalence of increasingly taller needlelike skyscrapers.  We have shown how the average height of buildings in metropolitan areas in North America can be predicted using urban scaling theory and how the shape and size of buildings can be characterized by one single scale and a set of shape parameters. As several dimensions of a building (width and height, for example) become similar, as we find in larger cities, the surface-to-volume ratio of buildings converges to its simplest case (analogous to a sphere). In such a case, or when shape parameters are statistically independent of building size, the allometric relation between building surface area and volume becomes simplest, $S_b \sim V_b^{2/3}$. This seems to be the limit to which buildings in North American cities converge to, on average, as cities become larger.  In this limit, thermal diffusion per person in buildings can be smallest (depending on other factors, such as thermal conductivity) and buildings can be most efficient in terms of energy management related to climate control. 

These results offer a quantitative baseline to predict the change in urban geometry as cities grow in population and wealth. This will be especially important in fast developing cities and may also find applications to shrinking urban areas, where such changes in urban form may happen in reverse. As the availability of building-related data grows quickly over the near future~\cite{Batty_2012}, it will be instructive to apply our framework to a larger number of cities in various cultures and economies. In a next step, to estimate city-wide energy consumption of buildings, our results can be combined with thermal models or directly applied in existing building energy simulators~\cite{Crawley_2001}. Besides urban energy and associated greenhouse gas emissions, it would be interesting to assess the impact of the changing building allometry on urban heat island effects~\cite{Zhao_2014} and other aspects of urban microclimates~\cite{Stewart_2014}. Finally, it remains an open challenge to combine city-wide building characteristics with large-scale data on the spatio-temporal distribution of people and their activities~\cite{Zhong_2015}, in order to better understand the complex interplay between socioeconomic processes and urban built form and to obtain improved systemic metrics of energy efficiency in cities.

\footnotesize
\section{Material and methods}

All data on building geometries used here were compiled from various publicly accessible sources. For New York City, the online portal {\it NYC Open Data} (http://nycopendata.socrata.com) provided us with detailed geospatial data on the building footprints (shapefiles containing spatial polygons), together with normalized building heights in terms of ground elevation and maximum height. We acquired the same type of data for Los Angeles (http://egis3.lacounty.gov/dataportal), Toronto (www1.toronto.ca), Boston (http://bostonopendata.boston.opendata.arcgis.com), Vancouver (http://vancouver.ca), Portland (www.civicapps.org), Austin (http://data.austintexas.gov), Ann Arbor (www.a2gov.org) and Santa Fe (www.santafenm.gov, upon request). These data were generated by the providing agencies using varying methodologies including the combination of LiDAR processing with orthoimagery and 3D-stereoscopy as well as cadastral and land assessor data. 

For San Francisco, we first sourced LiDAR data from the USGS Open Topography data portal (viewer.nationalmap.gov). From the classified LiDAR point cloud data, a digital elevation model (DEM) and a digital surface model (DSM) were generated. We then subtracted the DEM from the DSM yielding a normalized digital surface model (NDSM). Subsequently, the maximum height values from the NDSM were  attributed to each building footprint intersecting the NDSM. For this task, LASTools, a semi-opensource LiDAR processing library, was used in conjunction with GDAL, OGR, and Python~\cite{Isenberg_2015}. 

For Chicago, we extracted building heights from OpenStreetMap~\cite{Haklay_2008}. If absolute height values were not available, we used the number of floors and assumed a standard floor height of 3.5~m. If neither the absolute height nor the number of floors was given, we assumed the structure to be a single-story building. We did not extract building footprints.

For the city of Bellingham, besides the geographic location of the buildings, only the number of floors were provided (www.cob.org). To estimate the heights of buildings we used either information from OpenStreetMap or, if no height values were given, we multiplied the number of floors with a standard height of 3.5~m. For all cities studied here, we considered only structures that are taller than 2.5~m.  

Land and property values for single family homes in select US Metropolitan Statistical Areas were compiled by the Lincoln Institute of Land Policy and are publicly available at (www.lincolninst.edu).

\normalsize
\section*{Acknowledgment}

The authors thank Stephen Guerin and Carlo Ratti for helpful discussions. M.S. and L.M.A.B acknowledge financial support from the Army Research Office Minerva Programme (grant no. W911NF1210097). 



\newpage

\begin{table}[t!]
\caption{Summary statistics for building characteristics in 12 North American cities. The populations of the US cities (MSAs) are given by the 2010 US Census. For Canadian cities, the populations of Census Metropolitan Areas (CMAs, the equivalent definition to MSAs in the US), from the 2011 Canada Census are used. Building heights correspond to the year indicated, in which the raw data was collected, see Material and Methods. Maximum building heights for New York City have changed since 2013, so that the calculation of average building height does not include {\it One World Trade Center}, which opened in late 2014 and stands symbolically at 1776 feet (541m) tall.\vspace{0.4cm}}
\label{tab:statistics}
\resizebox{\columnwidth}{!}{
\begin{tabular}{| l | c | c | c | c | l |}
\hline
City & Population & Year & No. of  & Av. height & Max. height [m] \\[-0.2cm]
& & & buildings & [m]  & \\
\hline
New York City NY& 19,567,410 & 2013 & 1,066,354 & 8.4 & 377.6 \ Empire State Building \\
Los Angeles CA& 12,828,837 & 2008 & 1,071,512 & 5.6 &  309.9 \ U.S. Bank Tower\\
Chicago IL& 9,461,105 & 2015 & 811,359 & 4.8 &  442.0 \ Willis Tower \\
Toronto ON & 5,583,064 & 2015 & 389,985 & 5.6 & 443.0 \ CN Tower \\
Boston MA& 4,552,402 & 2012 & 80,409 & 9.0 &  240.2 \ John Hancock Tower \\
San Francisco CA& 4,335,391 & 2010 &151,787 & 9.7 & 259.2 \ Transamerica Pyramid\\
Vancouver BC& 2,313,328 & 2009 & 111,024  & 6.8 &  193.0 \ Living Shangri-la \\
Portland OR& 2,226,009 & 2014 & 580,103 & 4.7 &  163.1 \ Wells Fargo Center \\
Austin TX& 1,716,289 & 2013 & 452,439 & 7.0 &  210.3 \ The Austonian \\
Ann Arbor WI& 344,791  & 2009 & 35,054 & 5.6 &  85.3  \hspace{0.158cm} Tower Plaza \\
Bellingham WA & 201,140 & 2010 &  42,081  & 4.3 & 49.0 \hspace{0.158cm} Bellingham Towers \\
Santa Fe NM& 144,170 & 2006 & 29,498 & 4.1 &  25.7 \hspace{0.158cm} Eldorado Hotel\\
\hline
\end{tabular}
}
\end{table}

\newpage

\newpage

\newpage

\end{document}